\documentclass{ws-procs10x7-xxx}
\usepackage{balance}
\usepackage{epsfig}
\usepackage{graphicx}% Include figure files
\newcommand{\etal}{{\it et al.}}
\newcolumntype{d}[1]{D{.}{.}{#1}}

\makeindex
\begin{document}
\vspace{-0.5in}
\title{MEASUREMENT OF  $D_s^+\to\ell^+\nu$ AND THE DECAY CONSTANT
$f_{D_s}$}

\author{SHELDON STONE$^*$}

\address{Physics Department, Syracuse University, Syracuse, NY 13244, USA\\
$^*$E-mail: stone@physics.syr.edu}

\twocolumn[\maketitle\abstract{I report preliminary CLEO-c results
on purely leptonic decays of the $D_s$ using 195 pb$^{-1}$ of data
at 4.170 GeV. We measure $f_{D_s}=280.1\pm 11.6 \pm 6.0 {~\rm
MeV}$, and $f_{D_s^+}/{f_{D^+}=1.26\pm 0.11\pm 0.03}$. }
 \keywords{Leptonic decay; decay constant; charm decay. Date: October 9, 2006}]

\section{Introduction}
To extract precise information from $B-\overline{B}$ mixing
measurements the ratio of ``leptonic decay constants," $f_i$ for
$B_d$ and $B_s$ mesons must be well known.\cite{formula-mix} Indeed,
the recent measurement of $B_s^0-\overline{B}_s^0$ mixing by
CDF\cite{CDF} has pointed out the urgent need for precise numbers.
The $f_i$ have been calculated theoretically. The most promising of
these calculations are based on lattice-gauge theory that include
the light quark loops.\cite{Davies} In order to ensure that these
theories can adequately predict $f_{B_s}/f_{B_d}$ it is critical to
check the analogous ratio from charm decays $f_{D^+_s}/f_{D^+}$. We
have previously measured $f_{D^+}$.\cite{our-fDp,DptomunPRD} Here I
present the most precise measurements to date of $f_{D_s^+}$ and
$f_{D_s^+}/f_{D^+}$.

\begin{figure}[htbp]
 \vskip 0.00cm
 \centerline{ \epsfxsize=2.2in \epsffile{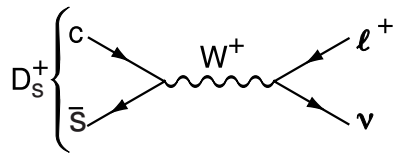} }
 \caption{The decay diagram for $D_s^+\to \ell^+\nu$.} \label{Dstomunu}
 \end{figure}

In the Standard Model (SM) the $D_s$ meson decays purely
leptonically, via annihilation through a virtual $W^+$, as shown in
Fig.~\ref{Dstomunu}. The decay width is given by\cite{Formula1}
\begin{eqnarray}
\Gamma(D_s^+\to \ell^+\nu) &=& {{G_F^2}\over
8\pi}f_{D_s^+}^2m_{\ell^+}^2M_{D_s^+}\\ [4pt]\label{eq:equ_rate}
&&\times\left(1-{m_{\ell^+}^2\over M_{D_s^+}^2}\right)^2
\left|V_{cs}\right|^2~, \nonumber
\end{eqnarray}
where $m_{\ell^+}$ and $M_{D_s^+}$ are the $\ell^+$ and $D_s^+$
masses,  $|V_{cs}|$ is a CKM element equal to 0.9737, and $G_F$ is
the Fermi constant.

New physics can affect the expected widths; any undiscovered charged
bosons would interfere with the SM $W^+$. These effects may be
difficult to ascertain, since they would simply change the value of
$f_i$ extracted using Eq.~(1). We can, however, measure the ratio of
decay rates to different leptons, and the predictions then are fixed
only by well-known masses. For example, for $\tau^+\nu$ to
$\mu^+\nu$:

\begin{equation}
R\equiv \frac{\Gamma(D\to \tau^+\nu)}{\Gamma(D\to \mu^+\nu)}=
{{m_{\tau^+}^2 \left(1-{m_{\tau^+}^2\over
M_{D}^2}\right)^2}\over{m_{\mu^+}^2 \left(1-{m_{\mu^+}^2\over
M_{D}^2}\right)^2}}~~. \label{eq:rat}
\end{equation}

Any deviation from this formula would be a manifestation of physics
beyond the SM. This could occur if any other charged intermediate
boson existed that coupled to leptons differently than mass-squared.
Then the couplings would be different for muons and $\tau$'s. This
would be a clear violation of lepton universality.\cite{Hewett}

\section{Experimental Method}
In this study we use 195 pb$^{-1}$ of data produced in $e^+e^-$
collisions using the Cornell Electron Storage Ring (CESR) and
recorded near 4.170 GeV. Here the cross-section for
$D_s^{*+}D_s^-$+$D_s^{+}D_s^{*-}$ is $\sim$1 nb. We fully
reconstruct one $D_s$ as a ``tag," and examine the properties of the
other. In this paper we designate the tag as a $D_s^-$ and examine
the leptonic decays of the $D_s^+$, though in reality we use both
charges. Track selection, particle identification, $\gamma$,
$\pi^0$, $K_S$ and muon selection cuts are identical to those
described in Artuso~\etal\cite{our-fDp}

The $D_s^-$ decay modes used for tagging are listed in
Table~\ref{tab:Ntags}. The number of signal and background events
are determined by fits to the invariant mass distributions.

\begin{table}
\tbl{Tagging modes and numbers of signal and background events,
within $\pm 2.5\sigma$ for all modes, except $\eta\rho^+$ ($\pm
2\sigma$), from two-Gaussian fits to the invariant mass plots.
\label{tab:Ntags}}
{\begin{tabular}{@{}lcccr@{}}\toprule
    Mode  &  Signal           &  Background \\ \colrule
$K^+K^-\pi^- $ & $8446\pm160$   & 6792\\
$K_S K^-$ & 1852$\pm$62 & 1021\\
$\eta\pi^-$; $\eta\to\gamma\gamma$ & $1101\pm80$  & 2803\\
$\eta'\pi^-$; $\eta'\to\pi^+\pi^-\eta$ & 786$ \pm $37  &242 \\
$\phi\rho^-$  & 1140$ \pm $59  &1515 \\
$\pi^+\pi^-\pi^-$ & 2214$ \pm $156  & 15668\\
$K^{*-}\overline{K}^{*0}$ & 1197$
\pm$81& 2955\\
$\eta\rho^-$  & 2449$ \pm $185  &13043 \\
\hline
Sum &  $19185\pm325 $ &44039 \\
 \botrule
\end{tabular}}
\end{table}

We search for three separate decay modes: $D_s^+\to\mu^+\nu$,
$D_s^+\to\tau^+\nu$, $\tau^+\to\pi^+\overline{\nu}$ and $\tau^+\to
e^+\nu \overline{\nu}$. For the first two analyses we require the
detection of the $\gamma$ from the $D_s^*\to\gamma D_s$ decay.
Regardless of whether or not the photon forms a $D_s^*$ with the
tag, for real $D_s^*D_s$ events, the missing mass squared recoiling
against the photon and the $D_s^-$ tag should peak at the $D_s^+$
mass and is given by
\begin{equation}
{\rm MM}^{*2}=\left(E_{\rm CM}-E_D-E_{\gamma}\right)^2-
\left(-\overrightarrow{p_D}-\overrightarrow{p_{\gamma}}\right)^2,\nonumber
\end{equation}
where $E_{\rm CM}$ is the center of mass energy, $E_{D}$
($\overrightarrow{p_D}$) and $E_{\gamma}$
($\overrightarrow{p_{\gamma}}$) are the energy (momentum) of the
fully reconstructed $D_s^-$ tag, and the additional photon. In
performing this calculation we use a kinematic fit that constrains
the decay products of the $D_s^-$  to the known $D_s$ mass and
conserves overall momentum and energy.

\begin{figure}[htb]
\vspace{-2mm} \centerline{\epsfig{file=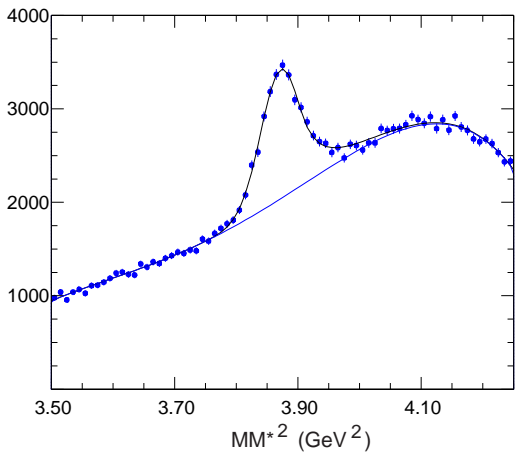,width=2.9in}}
\caption{The MM*$^2$ distribution from events with a photon in
addition to the $D_s^-$ tag. The curve is a fit to the Crystal Ball
function and a 5th order Chebychev background function.}
\label{MMstar2}
\end{figure}

The MM$^{*2}$ from the $D_s^-$ tag sample data is shown in
Fig.~\ref{MMstar2}.  Fitting shows a yield of 12604$\pm$423 signal
events. Restricting to the interval $3.978>$MM$^{*2}>3.776$ GeV$^2$,
we are left with 11880$\pm$399 events. The systematic error is
$\pm$4.3\%.

 Candidate $D_s^+\to\mu^+\nu$
events are searched for by selecting events with only a single extra
track with opposite sign of charge to the tag; we also require that
there not be an extra neutral energy cluster in excess of 300 MeV.
Since here we are searching for events where there is a single
missing neutrino, the missing mass squared, MM$^2$, evaluated by
taking into account the seen $\mu^+$, $D_s^-$, and the $\gamma$
should peak at zero, and is given by
\begin{eqnarray}
\label{eq:mm2} {\rm MM}^2&=&\left(E_{\rm
CM}-E_{D}-E_{\gamma}-E_{\mu}\right)^2\\\nonumber
         &&  -\left(-\overrightarrow{p_
         D}-\overrightarrow{p_{\gamma}}
           -\overrightarrow{p_{\mu}}\right)^2,
\end{eqnarray}
where $E_{\mu}$ ($\overrightarrow{p_{\mu}}$) is the energy
(momentum) of the candidate muon track.

We also make use of a set of kinematical constraints and fit the
MM$^2$ for each $\gamma$ candidate to two hypotheses one of which is
that the $D_s^-$ tag is the daughter of a $D_s^{*-}$ and the other
that the $D_s^{*+}$ decays into $\gamma D_s^+$, with the $D_s^+$
subsequently decaying into $\mu^+\nu$. The hypothesis with the
lowest $\chi^2$ is kept. If there is more than one $\gamma$
candidate in an event we choose only the lowest $\chi^2$ choice
among all the candidates and hypotheses.

The kinematical constraints are the total momentum and energy, the
energy of the either the $D_s^*$ or the $D_s$, the appropriate
$D_s^* - D_s$ mass difference and the invariant mass of the $D_s$
tag decay products.
 This gives us a total of 7 constraints. The
missing neutrino four-vector needs to be determined, so we are left
with a three-constraint fit. We preform a standard iterative fit
minimizing $\chi^2$. As we do not want to be subject to systematic
uncertainties that depend on understanding the absolute scale of the
errors, we do not make a $\chi^2$ cut, but simply choose the photon
and the decay sequence in each event with the minimum $\chi^2$.

We consider three separate cases: (i) the track deposits $<$~300
MeV in the calorimeter, characteristic of a non-interacting
$\pi^+$ or a $\mu^+$; (ii) the track deposits $>$~300 MeV in the
calorimeter, characteristic of an interacting $\pi^+$; (iii) the
track satisfies our $e^+$ selection criteria.\cite{our-fDp} Then
we separately study the MM$^2$ distributions for these three
cases. The separation between $\mu^+$ and $\pi^+$ is not unique.
Case (i) contains 99\% of the $\mu^+$ but also 60\% of the
$\pi^+$, while case (ii) includes 1\% of the $\mu^+$ and 40\% of
the $\pi^+$.\cite{DptomunPRD}

\begin{figure}[htbp]
 %\vskip 0.00cm
%\centerline{ \epsfxsize=3.0in
\centerline{ \epsfxsize=2.8in \epsffile{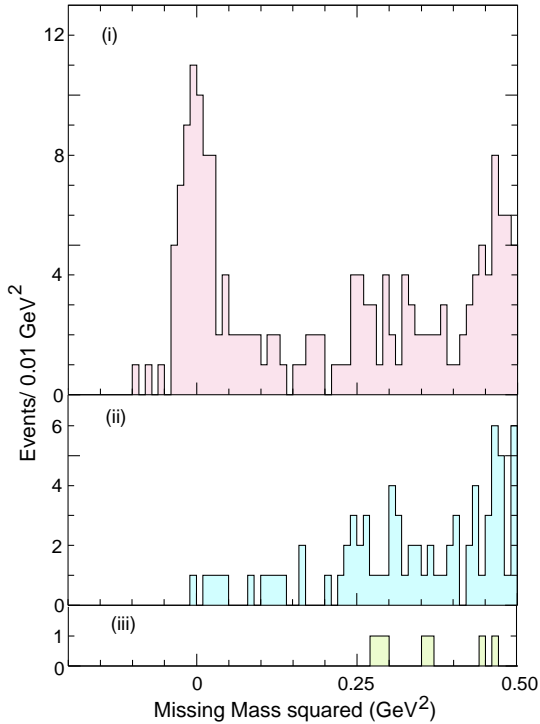}}
 \caption{The MM$^2$ distributions from data using $D_s^-$ tags and
 one additional opposite-sign
charged track and no extra energetic showers (see text). (a) Case
(i) the single track  deposits $<$~300 MeV of energy in the
calorimeter. The peak near zero is from $D_s^+\to\mu^+\nu$ events.
(b) Case (ii) Track deposits $>$~300 MeV in crystal calorimeter but
is not consistent with being an $e^+$. (c) Case (iii) the track is
identified as an $e^+$. } \label{mm2-data}
 \end{figure}

The overall signal region we consider is below MM$^2$ of 0.20
GeV$^2$. Otherwise we admit background from $\eta\pi^+$ and
$K^0\pi^+$ final states. There is a clear peak in
Fig.~\ref{mm2-data}(i), due to $D_s^+\to\mu^+\nu$. Furthermore,
the events in the region between $\mu^+\nu$ peak and 0.20 GeV$^2$
are dominantly due to the $\tau^+\nu$ decay.

The specific signal regions are defined as follows: for $\mu^+\nu$,
$0.05>$MM$^2>-0.05$ GeV$^2$, corresponding to $\pm 2\sigma$ or 95\%
of the signal; for $\tau\nu$, $\tau^+\to\pi^+\overline{\nu}$, in
case (i) $0.20>$MM$^2>0.05$ GeV$^2$ and in case (ii)
$0.20>$MM$^2>-0.05$ GeV$^2$. In these regions we find 64, 24 and 12
events, respectively. The corresponding backgrounds are estimated as
1, 2.5 and 1 event, respectively. The branching fractions are
summarized in Table~\ref{tab:results}. The absence of any detected
$e^+$ opposite to our tags allows us to set the upper limit listed
in Table~\ref{tab:results}.

\begin{table}
\tbl{Measured $D_s^+$ Branching Fractions \label{tab:results}}
{\begin{tabular}{@{}lc@{}}\toprule
    Final State  &  ${\cal{B}}$ (\%)       \\ \colrule
$\mu^+\nu$ & $0.657\pm 0.090\pm0.028$\\
$\mu^+\nu^{\dagger}$ & $0.664\pm 0.076\pm0.028$\\
$\tau^+\nu,~(\tau^+\to\pi^+\nu)$ & $7.1\pm 1.4\pm0.3$\\
$\tau^+\nu,~(\tau^+\to e^+\nu\bar{\nu})$ & $6.29\pm 0.78\pm0.52$\\
$\tau^+\nu$ (average) & $6.5\pm 0.8$\\
$e^+\nu$ & $< 3.1\times 10^{-4}$ (90\% cl)\\
 \botrule
\end{tabular}}
$\dagger$ From summing the  $\mu^+\nu$ and $\tau^+\nu$ contributions
for MM$^2$ $<$ 0.20 GeV$^2$.
\end{table}

\begin{figure}[htb]
%\centerline{ \epsfxsize=3.0in
%\centerline{ \epsfxsize=2.8in \epsffile{ecc-8.ps}}
\centerline{\psfig{file=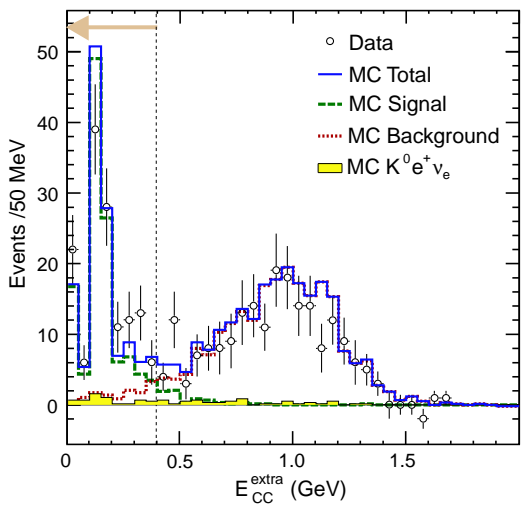,width=2.7in}}
 \caption{The extra calorimeter energy from data (points), compared with
 the Monte Carlo simulated estimates of semileptonic decays in general (dotted),
 the $K^0 e^+\nu$ mode specifically (shaded),
 as a sub-set of the semileptonics, and the expectation from signal (dashed).
The peak near 150 MeV is due to the $\gamma$ from $D_s^*\to\gamma
D_s$ decay. (The sum is also shown (line).) The arrow indicates the
selected signal region below 0.4 GeV.} \label{ecc-8}
 \end{figure}

The $D_s^+\to\tau^+\nu$, $\tau^+\to e^+\nu \overline{\nu}$ mode is
measured by detecting electrons of opposite sign to the tag in
events without any additional charged tracks, and determining the
unmatched energy in the crystal calorimeter (${\rm
E^{extra}_{CC}}$). This energy distribution is shown in
Fig.~\ref{ecc-8}. Requiring ${\rm E^{extra}_{CC}<}$ 400 MeV,
enhances the signal. The branching ratio resulting from this
analysis is listed in Table~\ref{tab:results}.

\section{Conclusions}
Lepton universality in the SM requires that the ratio $R$ from
Eq.~\ref{eq:rat} be equal to a value of 9.72. We measure
\begin{equation}
R\equiv \frac{\Gamma(D_s^+\to \tau^+\nu)}{\Gamma(D_s^+\to
\mu^+\nu)}= 9.9\pm 1.9~. \label{eq:tntomu2}
\end{equation}
Thus we see no deviation from the predicted value. Current results
on $D^+$ leptonic decays also show no deviations.\cite{ourDptotaunu}
Combining all three branching ratios determinations and using
$\tau_{D_s^+}$=0.49 ps to find the leptonic width, we find
 \begin{equation}
 f_{D_s}=280.1\pm 11.6 \pm 6.0 {~\rm MeV}.
 \end{equation}
Using our previous result\cite{our-fDp}
\begin{equation}
f_{D}^+=222.6\pm 16.7^{+2.8}_{-3.4}{\rm ~MeV,}
\end{equation}
provides a determination of
\begin{equation}
\displaystyle{{f_{D_s^+}}/{f_{D^+}}=1.26\pm 0.11\pm 0.03}.
\end{equation}

These preliminary results are consistent with most recent
theoretical models. As examples, unquenched lattice\cite{Lat:Milc}
predicts $1.24\pm0.01\pm0.07$, while one quenched lattice
calculation\cite{Lat:Taiwan} gives $1.13\pm0.03\pm0.05$, with other
groups having similar predictions.\cite{others}

\section*{Acknowledgments}
This work was supported by the National Science Foundation. I thank
Nabil Menaa for essential discussions.

\end{document}